# Fundamental hypotheses
# of the theory of elementary actions

Vincent Morin

Laboratoire Biostatistique et Informatique Médicales

22 Avenue Camille Desmoulins BP 815 29285 Brest Cedex

## Abstract

This text discusses the basic assumptions of a tentative unification theory called "Elementary Actions Theory". The final objective of this theory aims at solving the incorrectly called "quantum gravity" problem, and give a coherent view of todays physics.

Rather than trying at once a complete tour, we prefer to treat at length a single point. We will thus confine ourselves to the foundations, without discussing classical limits for now, leaving those for later discussions.

## A) Metaphysical basement.

### 1) Position of action as a physical quantity.

Action is a physical quantity whose importance is peculiar, since it is in Mechanics at the basis of Maupertuis, Lagrange et Hamilton, formulations, but also, and specially, because its quantification is the cornerstone of the Quantum Mechanics.

Examining the name of an old physical quantity is an instructive exercise. In effect, contrary to recent times, where exotic names are used with an implicit assumption that reality is out of reach of understanding, old physical names were thoroughly chosen to qualify a quantity in reason of its function in a physical explanation.

Change conceived as an alteration of status, supposes an action in the common sense. It is hardly conceivable that a system could change of status without any action being, even diffusively or subtiletedly, at its origin.

We say there is no alteration which does not result from an action. Action is the source of change.

Change is by nature discrete ; an infinitesimal change is the negation of itself because the contrast characteristic of change vanishes in the passage to the infimum limit.

That action, whose result is change, be quantified is not a surprise : we cannot believe that an infinitesimal action is at the direct origin of a discrete change.

Identification of the action physical quantity with that of the common sense proves interesting as we shall see.

The energy is a power of action, in the physical sense because its dimension is an action divided by a time, but also in a figured sense because acting with energy in effect signifies that actions are produced at a good rhythm and things evolve quickly.

Finally, if we are to build a physical theory of structures and their dynamics, choosing the action as a relational term between some fundamental elements is rather convincing.

Incidentally, consideration of energy as a power of action allows a diversification of the notion of form of energy : a physical system whose parts exchange lots of internal actions is the locus of an action dynamic whose power is the internal energy of the system.

This internal energy stays unchanged as long as no power of action is added or subtracted. The power of action or energy is conserved for the action dynamic of a stable isolated structure.

Exchange of energy is thus an exchange of dynamical structures of action, each having its own specific shape. Whence the same number of forms of energy.

We will see later that it is most probable that many systems are partly aspecific in their energetic exchanges, so that the precise form of energy exchanged is indifferent to them. An apparent reduction of the diversity in energy forms thus results.

The definition of a power of action implies the existence of a time to form the denominator of the quantity, we shall see later that even in absence of a time, a kind of variable density of actions replaces the power of action with a similar signification.

# B) Elementary actions theory postulates.

## 1) Postulates.

To conform to the preceding ideas, we shall build a model called « elementary actions model » and based on three postulates :

---

I. Substance postulate.

The physical universe is composed of fundamental elements (thereafter called "acteons") . Acteons are :

i. atomic in the original sense of the term,

ii. preexist to any geometrical arrangement,

iii. have a unique binary property (thus can have two states + or -),

iv. can change of state.

v. Acteons have a continuous existence where absolutely nothing intervenes between two successive state changes.

II. Co-relation postulate.

Acteons change their state by co-relation : an exclusive mutual action establishes, so that the state each acteon flips in the complementary state. An acteon cannot co-relates with itself.

III. Probabilistic dynamic postulate.

The dynamic development of co-relations obeys evolutive probabilistic laws. Those laws give at each universe alteration the probability for an acteon to co-relate with another different acteon.

---

## 2) Comments on Substance Postulate.

Acteons have a binary state and absolutely no other property (no mass, no locus, no charge, no wave function... nothing else than the binary property).

The binary property is not spin, nor charge, nor any physical quantity, but binary states will relate to charge polarity.

Acteons can take alternatively one of two states noted + or - ; each acteon has only one state in any segment of its existence, it never exists with both states. State alternance defines a succession of states, but not a time, each + state being absolutely identical to any other + state (the same for - states).

Acteons are not located in any space, nor exist in any time, but they are the basis for physics and space-time builds over those non-local and non temporal elements. Acteons are thus in a probabilistic sense over the entire universe (they can interact with any other acteon participating to any group if probabilistic dynamics allows).

Acteons have an existential continuity between two changes of state, though they are in absolute « immobility » (their only change is a binary state change).

This existential continuity has no other property than continuity being in absolute immobility (no time or distance or metric of any sort, nothing between state change).

The set of all acteons of the universe is noted E and contains N acteons :

$$E=\{a\}$$

$$Card(E)=N$$

N is supposedly a very great number, to give an idea, let us assume there is as many acteons in an atomic elementary particle than baryons in the universe, N then approaches $10^{160}$. Or else, let us assume there is as many acteons as Planck volumes in the universe's volume, N would then be around $10^{183}=(10^{26}/10^{-35})^3$. We will suppose N is between $10^{150}$ et $10^{200}$. This gives an idea of the hugeness of this number.

### 3) Comments on co-relation postulate.

Co-relation is a discrete mutual action a precursor of physical quantity action.

This mutual action is acausal and atemporal, whence the name co-relation (we shall sometimes write correlation).

To avoid confusion and useless periphrases, we call fundamental elements acteons because they act on each others.

The state transition of an acteon shall be noted $\rightarrow$ where necessary. Two transitions are possible which are noted 0 and 1 : $0=(+\rightarrow-)$ et $1=(-\rightarrow+)$.

We shall note $\leftrightarrow$ the co-relation of transitions. We shall call elementary action, or simply action if not confusing, the occurrence of this co-relation which produces the joint flips.

Four modalities of co-relations exist :

$$(0 \leftrightarrow 0), (0 \leftrightarrow 1), (1 \leftrightarrow 0), (1 \leftrightarrow 1)$$

### 4) Comments on probabilistic dynamic postulate.

We accept an irreducible ignorance of the final reason of elementary processes occurrence in the universe, and detailed development of this universe. Probabilistic dynamics embodies this view of unpredictability.

Simple structural hypotheses on dynamics randomness are necessary to reproduce physical events regularity and thus physical comprehensibility.

### 5) Final comments.

Those postulates give us a physical alterable substrate, a discrtete alteration mode resulting from a mutual action, and dynamics for those alterations.

We have the basis of a relational model of the physical universe with a minimum number of characteristics.

A good part of following text is devoted to stress what is not to be added to those postulates when representing it, and mathematically express the model for further treatment.

### 6) Short comments on contrasting with other theories.

Recent readings of some literature in the connex field of quantum gravity, deserve a few comments. The idea of discrete elements as a basis for a unification theory is not new, but details of concepts vary. R.Penrose with his spin networks, tried to build a geometry with spinning particle exchanges. Acteon are not particles, they do not have spin, the boolean property is not spin (they are not located in space, which does not exist, so do not rotate, they have no mass thus no angular momentum).

Some recent efforts of D.R.Finkelstein seem based on the idea of point swapping, acteons are not points.

Perhaps one would be tempted to call acteons qubits, I would discourage to use this name for several reasons. One is that scientists try to build quantum computers whose elements are called quantum bits or qubits ; this has evidently little in common with acteons. Another is that acteons have no quantum aspects (no wave particle dualism, absolutely nothing of the quantum theoretical formalism applies to them). The only fact that acteons have a binary property does not to call then quantum bits.

Some other discrete theories try other basis as causal sets (R.Sorkin). The acteons are not sets nor elements of causal sets, but we shall see that some causal structure perhaps related to Sorkin's causets does emerge of acteon dynamic.

## C) Representation and essential model properties.

It is possible to give a synthetic representation of a part of a physical universe evolution derived from the model, but many precautions are to be taken, so that no property is added to acteons. An example follows. The figure is just intended to suggest and discuss representation's irrelevant properties ; we shall precisely state in the technical section the correct building of such a graphic.

another + state, the same for − states.

Moreover nothing happens between two state transitions. So that we only have a cycle notion with no time.

4) The ordered sequence of successive states of an acteon is a sequence of existential states, so that each state segment of an acteon has an existence exclusive of that of the others state segments of the same acteon.

It also follows that actions between acteons have to form a kind of ladder to respect the existential order, because action can only take place between two states of common existence. If two acteons have been once co-related, no co-relation can occur between a following state of the first acteon and a preceding state of the second : such a co-relation would join two states with no common existence.

## 1) Analysis of a model universe.

If we agree with the model above, there is a particular arrangement amongst all possible, which represents the real universe.

The knowledge and thus the analysis of this particular universe is out of reach, because of our limited means and impossible detailed knowledge of the development. We are thus forced to a probabilistic approach.

Let us examine how to build a model universe in mathematical terms.

To each acteon $a_i$ in each state of its existence we attach a probability measure derived from a stochastic kernel $\mu_i = N(a_i, \ )$. So that $\mu_i(a_j)$ is the probability for acteon $a_i$ to correlate with $a_j$. The kernel can change at each step of the universe building.

We can mechanically construct a universe by drawing at each step an N by N graph adjacency matrix of null square in the set $A$ of all such matrices. Such a matrix describes a graph on vertices acteons made of chains of length one. Each step gives us an at once choice of a number of co-relations. The probability $Q_t(A)$ of a matrix $A$ occurrence at step $t$ can be calculated from $N_t$, the stochastic kernel for acteons co-relations.

We in fact define a stochastic process in the Doob sense :

$$P = (U, \mho, P, A, \mathbf{N}, (X_t)).$$

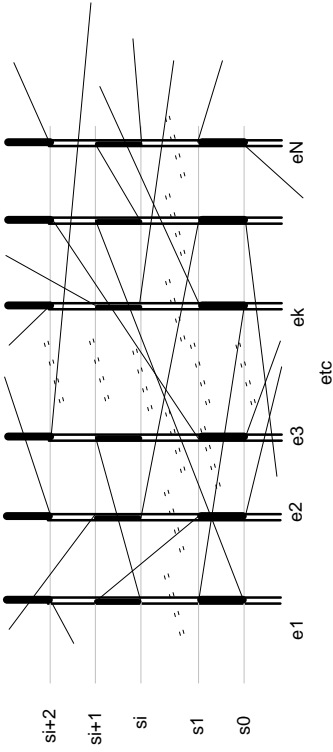

This representation calls for several series of important comments.

1) Concerning the signification of each component of the figure : we indexed each acteon from $a_1$ to $a_N$ (indexation is of course conventional). The ordered states of each acteon are represented with a vertical bar wich signifies the existential continuity, and the black and white segments represent + and − states respectively. States have been indexed arbitrarily and every other indexation (choice of « coordinates ») is equally valid.

Thin lines joining two transitions indicate the co-relations.

Some dashed lines just suggest the elements which have not been represented.

2) The fact that we are drawing on a paper sheet imposes constraints which do not belong to the model :

The length of black and white state segments has no signification, they are taken equal for convenience of indexing. The same remark consequently holds for thin action lines.

Localisation is wrongly suggested by vertical existence bars : acteons have no size and no locus, they « float nowhere, in nothing with a non-existent size ».

Briefly stated, nothing more than what is stated in the model's postulates is to be considered, everything else is artifact of representation.

3) The verticality of acteon existence bars and the progression of state changes wrongly suggest a notion of time. There is an ordinality, but no time evolution, because for each acteon every + state is strictly identical to simplest initial state for the universe : all acteons probability measures are uniform distributions. Each acteon thus has the same probability to co-relate with any other, there is no structural differentiation. As we will see that separations can be defined through co-relation bunches, this equiprobable uniform state of the universe puts everything at a same small separation.

This initial state is highly improbable, and negligibly entropic.

The only possible evolution is an toward differentiation and structural organisational complication. In a represention space for acteons co-relations probability measures, the state of the univers becomes more entropic. In the same time, separations between emerging structures increase, and the universe expands.

We thus have a simple explanation of the universe inflation, though it appears in the acteon model as a "static" inflation relatively to the set of non local acteons, only resulting from relational differentiation and increasing organisational complexity in co-relation probabilities. More detail in the following will enlight this view.

## D) Relational definition of structures

We can think action relations are more frequent (in statistical sense) between some acteons, rarer between others. Whence the idea that structures are distinguished by an aspect of inhomogeneity in action frequencies between elements belonging (perhaps transiently) to the structure and others outside the structure. But this simple idea analyses in more complex aspects as the following shows.

### 1) Simple homogeneous structures

First we can express an homogeneous structure notion by associating to each acteon those with which it has greater probability of interacting. The set of acteons belonging to such a structure are those which have greater interaction probability between them. This define an internally coherent structure. By qualifying of homogeneous this kind of structure, we assume that co-relations probabilities between structure acteons have sensibly identical flat profiles. We will adhere to this natural definition of a simple homogeneous structure.

To express this, we shall say that such a structure S is a set of $N_S$ acteons

The base space U is the set of all possible universes, U the σ-algebra on U. P a probability measure on U giving the probability of a specific universe. The state space is A the set of all N×N adjacency matrix of null square, the process time is A (do not confuse with a physical time ! even if t is used as index) is the set of naturals, and all universes structures are contained in stochastic variables $X_t$.

Though very compact and widely used in the mathematical litterature, this process definition is in fact very abstract and has been criticised for its somewhat counter intuitive character by P.Levy.

The equivalent formulation, which is probably more familiar to physicists, is in term of successive transition probability laws, and uses the same process time in N and state space A. But this formulation puts in prominent place the laws $Q_t$ (A) on A, derived from kernels $N_t$. The abstract character of the base space U and its probability law P with subsumes all transitions of all universes through random variables $X_t$, is absent.

We will prefer, wherever possible the description in term of kernels $N_t$ at each step.

Some simplifications can be easily described. For example one may want to use only one kernel N for all process times (or at least a great number of process steps, assuming a quasi stability of co-relations probabilities). In fact, each state of an acteon $a_i$ can have a very different co-relation probability, we thus have to take two kernels $N^+$ and $N^-$ for all steps t considered, where $N^+(a_i;)$ gives the co-relation probabilities of $a_i$ when it is + (symetrically for $N^-$).

### 2) The representation space M for acteon measures

Let ($a_i$) be the family of N acteons, there is a vectorial space M with N real dimensions, where a point $P_i$ represents the probability measure $\mu_i$ for co-relation of an acteon $a_i$ with other acteons. This space will be useful to describe the dynamics of probability measures.

### 3) The starting state and universe expansion.

The probabilistic description of the acteons relations, suggests the relation probability) and approaches from other structures (increases co-relation probability).

## E) Global relational characterization with action bunches.

A structure has a changing state consequently to action co-relations with its outside. Some of those actions, taken in the universe building sequence, conceptually form a bunch whose detailed sequence and arrangement can presumably, in some circumstances, have a determinant effect ; but we shall suppose that in many cases it is reasonable to neglect the specific arrangement of those external actions, and that a global qualification of an action co-relations bunch is appropriate. We shall see below how we attach a complex number to an action co-relations bunch.

### 1) Structure relational closing, inertia and mass

Internally coherent structures are composed of acteons which have the tendency to co-relate more with acteons inside the structure than with acteons outside. A structure thus shows a kind of closing on itself relatively to the external environment. Said in other terms, such a structure has internal action co-relations bunch denser than co-relations bunches with other external structures. We will also say a structure is more or less isolated from its environment, so that many internal co-relations are produced when only a few co-relations with the outside exist.

From this remark, we can infer that the more a structure is closed on itself, harder it will be to affect it (if no internal amplifying mechanism or sensitive point effet operates in the structure, this will be assumed in the following).

We can feel a closed structure will show a generalized inertia effect (difficulty to affect) in proportion of its isolation. The special case of mechanic alteration of movement (acceleration) with difficulty proportional to mass, will be so related to structural closing. We shall see that geometry being defined by inter structure co-relations bunches, geometry will be affected by structure isolation (mass-energy).

$a_j$ which have $\mu_i(a_j)=k_d$ if $a_j$ is in S and $\mu_i(a_j)=\mu_{ij} < k_d$ if it is outside S.

For S to be sufficiently distinct from the outside, $\mu_{ij}$ is supposed well below $k_d$.

One consequence of this structure relational scheme is that much more co-relations will occur homogeneously between all acteons inside the structure, and much fewer between the structure and the outside.

### 2) Externally coherent structure.

But another expression could be imagined by associating acteons on the basis of the similarity of their profile of co-relation probabilities. That is two acteons are grouped because they interact similarly with the outside, independently of their direct co-relation probability. This defines a kind of externally coherent structure.

To this end we define a scalar product in the representation space M for the co-relation probability measures of acteons a and b, $\mu_a$ and $\mu_b$:

$$\langle \mu_a, \mu_b \rangle = \sum_i \mu_a(a_i) \mu_b(a_i) = E_a(X_b) = E_b(X_a)$$

For an acteon a, we can define a neighbourhood $B_p$ which contains all acteons $a_j$ whose $\mu_j$ has a scalar product with $\mu_a$ which exceeds p :

$$\forall a \in E, B_p(a) = \{a_j \perp \langle \mu_i, \mu_a \rangle > p\}$$

External coherence is not exclusive of internal coherence as defined just before. We can feel that an internally coherent structure has a defined itself with its internal co-relations and is in a sense externally localized if an external coherence also exists (and in a sense dispersed if not).

### 3) Relational places and processual distances.

Those conceptions of structure imply an intuitive definition of a relational form of place. With the internal coherence definition, two structures whose acteons have great probabilities to co-relate are closer (share a place) than two which weakly inter co-relate. Also a processual definition of distances follows as alteration of the relational state of a structure which increases distance with some structures (decreases co-

We call isolation of an acteon subset S the ratio $I=Ni/Nx= t_i / t_x$. S is unisolated if I is zero, totally isolated if $I=+\infty$.

An acteon is totally extraverted and unisolated, the universe is totally intraverted and isolated. Between those extremes, imbricated structures do intravert with composition scale climbing, the rythm of intraversion depending fom co-relation probabilities.

From the preceding, we infer that, ascending from acteon level to universe level, there is a structural composition scale where structures do intravert. Conversely, seen from the macroscale with the structural decomposition point of view, intraverted macro structures do extravert at some scale when descending in the microscale. It is the notion of structural retroversion which was pointed out in an old article[1].

An acteon set with great intraset co-relation probabilities intraverts with fewer acteons than another acteon grouping weakly coupled in probability.

To take an example, let us assume that on $N=10^{160}$ acteons in the universe, $10^{80}$ of them are in a particle structure S and have an intraset co-relation probability very close to $10^{-80}$. This subset is globally nearly totally intraverted because each acteon of the structure will nearly always co-relate inside, but each part of S is less intraverted, and half of S has only a null intraversion (each of its acteons has the same chance to co-relate in the considered half or outside, that is the other half).

The cohesive strength of a structure is represented by the value of its internal co-relation probabilities.

The retroversion scale can be defined as the decomposition scale level where, cohesive probabilities being taken in account, the components structures at the considered level extravert. The more probabilistic cohesion we have in a structural composition, the deeper the level where extraversion occurs.

It is a reason why we suspect that Planck length does not exists as a common reality, but only as the localization limit conceptually obtainable if sufficient energy is involved in the tentative to localize a single acteon. But

## 2) Co-relation bunches balance

Let us consider two acteons subsets $S_1$ et $S_2$ of the whole acteons set E. Then consider global action exchanges in a part of the universe building process with five bunches. The following figure illustrates the situation :

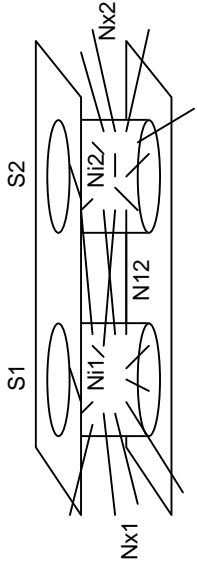

There is a total $T_1$ of co-relations concerning $S_1$, and $T_2$ for $S_2$.

$T_1$ is decomposed in $Ni_1$ actions exchanged properly inside $S_1$, $Nx_1$ actions exchanged with the outside amongst them $N_{12}$ actions are exchanged with $S_2$. We have the same for $S_2$, with $Ni_2$, $Nx_2$ et $N_{12}$.

In the following, we note $N_1=Ni_1+Nx_1-N_{12}$ (number of actions concerning $S_1$ omitting those relating to $S_2$) and the same $N_2$ for $S_2$.

Unless otherwise noted in the following we assume that $Nx_1$, $Nx_2$ and $N_{12}$ are well below $Ni_1$ and $Ni_2$. (we simply assume that structures are internally coherent and to some extent isolated).

Some extreme cases can be examined. If $N_{12}$ is zero, $S_1$ is totally isolated from $S_2$ (and conversely). If $Nx_1$ is zero, $S_1$ is totally isolated from its environment (we assume it never happens on the whole universe development, in which case it would not be connex).

If $Ni_1$ is zero, $S_1$ has no cohesion, $Nx_1$ being non zero, $S_1$ is wholly externally connected. If then $Nx_1=N_{12}$, $S_1$ is completely $S_2$ related.

Let $t_x=Nx/T$ the external relation ratio, and $t_i=Ni/T$ the internal relation ratio.

We call intraversion of an acteon subset S the ratio $(Ni-Nx)/T= t_i - t_x$. A totally extraverted acteon set (Ni=0) has intraversion -1. A totally intraverted (Nx=0) has intraversion 1.

---

[1] V.Morin Bio-Math Tome XXXII 3iè trimestre 1994

we think that this extreme situation is probably not a natural one, but only the ultimate limit of our experimental analytical approach which analyses structures in local contiguous parts (super microscopic approach). It is not sure that nature commonly relates structures as our experimental setups do, we tend to believe the contrary (that nature does not analyses structures as we do).

This retroversion phenomenon should be finely analysed specially in relation with baryon quark substucture and confinment phenomena, but that will not be attempted for now ; we shall first concentrate on geometry and gravitation which is the scene for electrodynamics and quantum mechanics.

### 3) Complex number description of action bunches.

To globally qualify an action co-relations bunch, we shall associate to each of the four co-relation modalities a unimodular complex number as follows :

$$(0\leftrightarrow 0)=1, \ (1\leftrightarrow 0)=j, \ (1\leftrightarrow 1)=-1, \ (0\leftrightarrow 1)=-j$$

Each action co-relations bunch will consequently be represented by a complex number $n_1+j\, n_2$. Here $n_1$ represents the balance between $(0\leftrightarrow 0)$ and $(1\leftrightarrow 1)$ co-relations, and $n_2$ the balance between $(1\leftrightarrow 0)$ and $(0\leftrightarrow 1)$.

We thus have the possibility to represent an alteration (state change) of a structure with omission of transformation details. This representation method can be used for a particle state to give the transformation from a référence state to each state of the particle's stability state set. But we shall first use this representation to define geometry through spinors.

Let us detail further. To a bunch of N co-relations which contains $N_{00}$, $N_{10}$, $N_{11}$, et $N_{01}$ co-relations of each modality indicated in index, we associate the complex z :

$$z = D_s + j D_a = (N_{00} - N_{11}) + j(N_{10} - N_{01})$$

Incidentally, let us note that multiplying each co-relation modality by j gives the following correponding to the application of an operation R which binary complements the second transition and permutes with the first :

$$R(0\leftrightarrow 0)=(1\leftrightarrow 0) \quad R(1\leftrightarrow 0)=(1\leftrightarrow 1) \quad R(1\leftrightarrow 1)=(0\leftrightarrow 1) \quad R(0\leftrightarrow 1)=(0\leftrightarrow 0)$$

Each of the N co-relations counts for a little unimodular complex, and drawing them one at the end of the other produces a random walk on a unit square grid in the complex plane **C** :

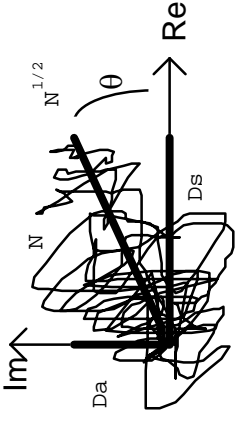

Representation of a random walk generated by N co-relations
The displacement is in square of N, Ds et Da being coordinates of displ.

If this random walk is a brownian motion we have :

$$D_s = (N_{00} - N_{11}) = k_s\sqrt{N} \quad \text{et} \quad D_a = (N_{10} - N_{01}) = k_a\sqrt{N}$$

Thus an action co-relations bunch is qualified by the complex :

$$z = \sqrt{N} e^{j\theta}$$

The application of the R operation (j multiplication) to randomly chosen unimodular segments of the random walk induces its rotation, so that the angle θ can be interpreted in term of a number of R operations.

We can reconsider the two structures' relationship and give more precise qualification to the five action co-relations bunches involved with $S_1$ and $S_2$. Each of the five buches is described by a complex : $z_1$ (inside bunch of $S_1$), $z_2$ (inside bunch of $S_2$), $z_{12}$ (inter bunch between $S_1$ and $S_2$), $z_{x1}$ ($S_1$ external bunch, omitting inter), $z_{x2}$ ($S_2$ external bunch, omitting inter). This being schematized below :

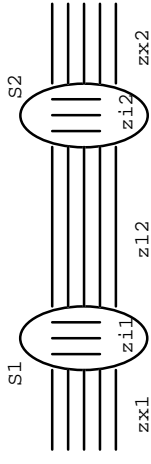

We can then qualify the relation of $S_1$ with $S_2$ through the inter bunch by writing a complex two component vector:

$$\eta^A = \frac{1}{Z_{12}}\begin{bmatrix} z_{i1}+z_{x1} \\ z_{i2}+z_{x2} \end{bmatrix} = \frac{1}{\sqrt{N_{12}}}\begin{bmatrix} \sqrt{N_1}e^{j(\theta_1-\theta_{12})} \\ \sqrt{N_2}e^{j(\theta_2-\theta_{12})} \end{bmatrix}$$

This expresses the disposition of non inter bunches relatively to inter bunch by two complex ratios which give two non inter complex qualifier of related structures when multiplied by inter bunch complex qualifier.

## 4) Spinorial interpretation of inter structure action relation.

A possible technical transition technique from probabilistic co-relations bunches to mean geometric behaviour is through the identification of this inter structure complex vector with a two component spinor.

This identification agrees with the spinor rotation-entanglement interpretation which has already been put forward in[2].

If we proceed further, a null four vector is attached to the two structures with the classical "null pole" formula:

$$u^\alpha = \begin{bmatrix} u^t \\ u^x \\ u^y \\ u^z \end{bmatrix} = (u^\alpha)^{AA'} = \begin{bmatrix} u^t+u^z & u^x+ju^y \\ u^x-ju^y & u^t-u^z \end{bmatrix} = \eta^A \eta^{*A'} = \begin{bmatrix} \eta^1\eta^{1*} & \eta^1\eta^{2*} \\ \eta^2\eta^{1*} & \eta^2\eta^{2*} \end{bmatrix}$$

From the preceeding we can write the null separation in term of co-relation numbers:

$$u^t = \frac{1}{2}\frac{N_1+N_2}{N_{12}}$$
$$u^z = \frac{1}{2}\frac{N_1-N_2}{N_{12}}$$
$$u^x = \frac{\sqrt{N_1N_2}}{N_{12}}\cos(\theta_1-\theta_2)$$
$$u^y = \frac{\sqrt{N_1N_2}}{N_{12}}\sin(\theta_1-\theta_2)$$

By this procedure, the action co-relations bunches which concern the two structures $S_1$ and $S_2$ define a null separation characteristic of a situation in a light cone. This could be taken as a confirmation of the fundamental importance of the null structure of space-time.

## 5) Time separation.

In the above formula, the time separation $u^t$ is only positive and attached to a spatial separation, there is no natural mean to give a sign to this time separation and detach it from its spatial counterpart.

But for a given set of acteons we can define a signed time separation, and thus a proper time with the following scheme:

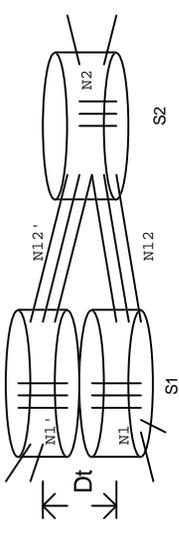

Two séparations $u^t$ et $u^t$' can be defined between a first state of $S_1$ and a second, relatively to a state of $S_2$. So that we write a Dt by difference:

$$Dt = \frac{1}{2}\left(\frac{N_{1'}+N_2}{N_{12'}} - \frac{N_1+N_2}{N_{12}}\right)$$

Supposing $N_{12}' = N_{12} + \varepsilon$ et $N_1' = N_1 - \varepsilon$ with $\varepsilon << N_{12}$:

$$Dt = -\varepsilon\frac{N_1+N_{12}+N_2}{2N_{12}(\varepsilon+N_{12})} \approx -\varepsilon\frac{N_1+N_{12}+N_2}{2N_{12}^2}$$

---

[2] Misner Thorne K.S. Wheeler J.A. "Gravitation", Freeman, 1972.

This definition is more complex than it appears at first, because no restrictions have been put on the external coherence of structures involved (spatial incoherence of time, or non local subjective time), and nothing has been discussed concerning the problem of common existence of acteons states related by the inter bunches (question about the thickness of structure existence).

## F) Geometry from structure inter-relations.

### 1) The pseudo cristal of test particles.

To resolve the problem just mentioned above and place ourselves in a simplified context for geometrodynamics, we shall further assume that all structures are identical test particles such that probabilities of inter-structures co-relations have a specific form that defines a locus for each test particle.

The test particle is defined as a set of acteons forming a well intraverted homogeneous structure.

To determine those inter structure co-relation probabilities, we can take a model 3D space M (supposed sufficiently simple to avoid multiple paths between two points). Then we define a one to one application of an evenly chosen set of points of M to the set of all acteons (we associate a point to an acteon, but recall an acteon is *not* a point, the application is just a tool to calculate probabilities). Each $\mu_{ij}$ value is in proportion of $1/d$ the inverse geodesic distance between points associated to acteons $i$ and $j$. The model space M can then be forgotten.

Then acteons are grouped with a sufficient number of their neighbours to form intraverted groups, the test particles. We thus end with a set of structures which form a pseudo cristal in a statistical sense because separations are defined by co-relations probabilities between non local acteons, but such that inter structure separations reproduce the distances of the model space (the $N_{12}$ factor of the spinorial approach is responsible of distance elaboration).

### 2) Metric from test particles.

In the context exposed above, we build the geometry of space-time by a process inspired from the Regge calculus and through the "relation to spinor to null separation" and "time separation" definitions as given before.

Given 5 test particles states o,a,b,c,d in space-time, o being taken as an origin point for independant basis vectors $\mathbf{e}_\alpha$ between o and the four other points. The metric coefficients expressed in squared length between points are :

$$g_{\alpha\beta}=\frac{-1}{2}\begin{bmatrix} 2l_{oa}^2 & l_{oa}^2+l_{ob}^2-l_{ab}^2 & l_{oa}^2+l_{oc}^2-l_{ac}^2 & l_{oa}^2+l_{od}^2-l_{ad}^2 \\ & 2l_{ob}^2 & l_{ob}^2+l_{oc}^2-l_{bc}^2 & l_{ob}^2+l_{od}^2-l_{bd}^2 \\ & & 2l_{oc}^2 & l_{oc}^2+l_{od}^2-l_{cd}^2 \\ & & & 2l_{od}^2 \end{bmatrix}$$

If three points b,c,d are on the past null cone of the first point o taken as origin and the fifth point a gives $\mathbf{a}=\mathbf{e}_0$ with no spatial component (pure time), we have a simplified metric tensor :

$$g_{\alpha\beta}=\frac{-1}{2}\begin{bmatrix} 0 & l_{ab}^2 & l_{ac}^2 & l_{ad}^2 \\ & 0 & l_{bc}^2 & l_{bd}^2 \\ & & 0 & l_{cd}^2 \\ & & & 0 \end{bmatrix}$$

If o is taken as origin, a,b,c have coordinates given by their spinors with o, for example $\mathbf{b}$ is such that :

$$\mathbf{b}^0=\frac{1}{2}\frac{N_o+N_b}{N_{ob}} \quad \mathbf{b}^3=\frac{1}{2}\frac{N_o-N_b}{N_{ob}} \quad \mathbf{b}^1=\frac{\sqrt{N_oN_b}}{N_{ob}}\cos(\theta_o-\theta_b) \quad \mathbf{b}^2=\frac{\sqrt{N_oN_b}}{N_{ob}}$$

But :

$$l_{ab}^2=-(\mathbf{b}^0-\mathbf{a}^0)^2+\sum_{i=1}^{3}(\mathbf{b}^i-\mathbf{a}^i)^2$$

$$\mathbf{a}^0=\frac{1}{3}\sum_{i=b,c,d}\frac{1}{2}\left(\frac{N_o+N_i}{N_{oi'}}-\frac{N_o+N_i}{N_{oi}}\right)$$

The metric is thus expressed with $N_o$, $N_b$, $N_c$, $N_d$, $N_o'$, $N_{ob}$, $N_{oc}$, $N_{od}$, $\theta_o$, $\theta_b$, $\theta_c$, $\theta_{d}$..

process of relational alteration, that is the process through which a structure passes to transform its relations defining the initial locus to relations defining its final locus (displacement of the representative cloud of points in the space of co-relation probability measures).

The richness of those definitions of locus and distances is far greater that the common behaviour of our space ; this posing a question to which we shall give an answer by telling that our common space behaviour is an average behaviour.

In a universe with structures as we defined them above (an intraverted set of acteons), each structure S has a generalized neighbourhood (not necessarily localized) defined as the set of structures with which S is in greater relation. Neighbourhoods of increasing extension are build by addition of structures with ever weaker relation.

The transfer of a structure from a neighbourhood to another demands a redefinition of co-relation probabilities through a process of state transformation. This corresponding to a generalized displacement of S (not necessarily linear, nor local or continous in a space defined as the pseudo-cristal of test particles.

The quantum wave packet reduction and the uniform relativistic movement are two such processes ; the first case being non local.

## 5) Evolution of separation, inertia and energy-momentum.

From the definition of separation between two acteon sets (with a spinor representing the action co-relation bunch), we can extract an evolution of separation which translates into conventional speed for localized and intraverted structures. Taken in account for isolation, impulsion can also be derived.

Let us consider a "z" separation, it depends on two factors $N_1$-$N_2$ at numerator, $N_{12}$ at denominator ; it thus can vay because the difference of intraset action numbers vary or because the interset relation level changes. If internal relational conditions do not change, $N_{12}$ variation is responsible for separation variation.

## 3) Mass and space-time geometry.

Mass has been previously related to isolation and the point is the statistical mean locus of a test particle. A simple argument can be given to explain why curvature is generated by isolation, this reinforcing the idea of isolation related to mass-energy.

The following drawing illustrates what happens. If all test particles have an identical isolation, 1/r relations define statistically identical action bunches between particles similarly disposed.

If isolation of one of the test particles is increased, relations from close test particles arrive in lesser number on the isolated particle and dispatch on other test particles in the environment. Peripheral relations are thus increased as radial relations decrease. (the radial $N_{12}$'s decrease, at constant radius the $N_{12}$'s increase). Peripheral separations decrease as radial separation increase.

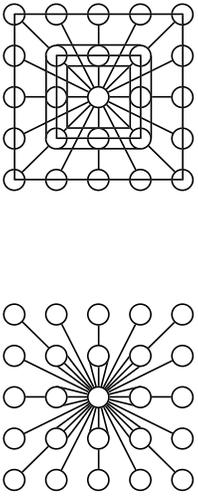

But this phenomenon is characteristic of curvature, perimeter to radius ratio being less than $2\pi$.

If relations rarefy in space, they also rarefy in the universe development. But existence continuity lines, grouped to define a mean existence line for the test parcicle, form a timelike line. A space-time curvature thus originates from isolation.

## 4) Relational definition of locus and processual definition of distances

Preceding discussions associate the notion of locus with that of relation. We shall distinguish the notion of static separation as action bunches define it, and distance as a separation to pass over. Distance is then defined as a

We also shall say that the electron is an electric structure, to have a generic term applicable to cases other than the negaton/positon association (for other particles).

Negaton and positon have, in our view, the possibility to be extremely non local in the sense that their constitutive acteons can interact with intraverted structures largely dispersed in a constituted physical space as we experience it. Quantum mechanics seems to say something comparable, but in fact this acteon set dispersion and consequent non locality is of a nature different from wave function spreading.

It is a natural to assume that a charged structure is an acteon set which maintains a vast majority of + states (or - states), the acteon polarity is then used for charge polarity without having to invent another characteristic. This does not imply that an acteon is a charge as under the term of charge there is a whole set of implied characteristics from classical electromagnetism (for example the $1/r^2$ Coulomb interaction) which acteon do not have. (recall that no other characteristic other than those of the basic hypotheses must be added for acteons).

But this simple use of acteon polarity for charge polarity brings in a difficulty : how can the specific polarity state be generally maintained as acteons change of state at each co-relation ? We are forced to assume that while acteons have the charge polarity, they interact with the outside acteons as required to define the charge locus ; but as soon as they flip their state, they interact with the outside acteons in quite another way, without ceasing to interact among them (in the electric structure) the same way.

Doing this, we do not create a new basic hypothesis, but only give a special polarity dependence to some acteon co-relation probabilities.

What we say there, is that a charge (say the negaton) is only half of an electric uncharged structure (the redefined electron) which results from the term electron for another use), its antiparticle "positon" as is usual, and, beware of this, "electron" is redefined as the association of a negaton and a positon in one coherent intraverted structure of acteons. The other half (the positon) being constituted of the oppositely signed acteons of the electron and which is apparently invisible just because it is extremely dispersed (we pointed out this possibility at the beginning of the paragraph).

If we take notations and conventions of the preceding paragraph for Dt, separation and relative z speed are :

$$Dz = -\varepsilon \frac{N_1 + N_{12} - N_2}{2N_{12}(\varepsilon + N_{12})} \approx -\varepsilon \frac{N_1 + N_{12} - N_2}{2N_{12}^2} \quad v_z = \frac{Dz}{Dt} = \frac{N_1 + N_{12} - N_2}{N_1 + N_{12} + N_2}$$

We see that relative speed of $S_1$ against a state of $S_2$ is necessarily upper limited to 1 in absolute value, and that more singularly, $\varepsilon$ has no effect whatever its value. Action number random fluctuations produce an $\varepsilon$ and thus a Dz and a Dt, so that $V_z$ never corresponds to a 0/0 ratio.

An acteon set $S_1$ isolated from another $S_2$ is inert to change induced by $S_2$. This is true for sufficiently important action bunches though, because at extreme limit, a lone acteon, which is totally extraverted (thus not isolated at all), is so fluctuating in its relations that no ordered action can exist.

In fact inertia of $S_1$ for $S_2$ is defined as difficulty for $S_2$ to provoke an ordered change of $S_1$. An acteon can thus appear as a very inert entity for the acteon sets that our apparatuses and us are ; but the non local nature of this acteon explains the absence of conventional curvature generation.

If isolation $2N_1/(N_{12}+N_{12'}) \approx N_1/N_{12}$ is taken as mass (to a factor of unity for dimension), we can write momentum of $S_1$ seen by $S_2$ (in the frame given by $S_2$ or relatively to $S_2$) :

$$E_{S_1} \approx \frac{N_1}{N_{12}} \quad p_z = m_{S_1} v_z \approx \frac{N_1}{N_{12}} \frac{N_1 + N_{12} - N_2}{N_1 + N_{12} + N_2}$$

## 6) Electric structure and charge.

Do take important note of this (re)definition of the electron term : to simplify the terminology we shall call the conventional electron "negaton" (to free the term electron for another use), its antiparticle "positon" as is usual, and, beware of this, "electron" is redefined as the association of a negaton and a positon in one coherent intraverted structure of acteons. The justification to redefine the electron as an uncharged association to designate an acteon structure fundamental to electrodynamics follows.

### a. Specificitie of charge structure action bunches

An electric structure has internal actions globally described by an internal bunch and a descriptive spinor (as every structure has). The preceding figure separates the internal action bunch in three internal action sub-bunches for an electric structure, one of them between charged poles globally noted $z_I$. The other two are intra polar. In the special case where the internal bunches contain only co-relations which reverse only once each acteon (co-relations of a reversal bunch), the two extremities of the $z_1$ bunch divide the electric structure acteon set in two halves (this is not the case anymore when several reversals are in the bunch because acteons alternatively belong to each pole at each reversal, so that there is generally several co-relation links for a same acteon successively in both poles when several reversals are bunched).

Let us assume that an electric structure Q be constituted with K acteons forming an intraverted set. Close to K/2 acteons have polarity -, belong to the minus pole and have similar $\mu_i(a_i)$ measures as defined by the scalar product of paragraph D.2 with barycenter G-. Other K/2 acteons are +, similar among them with a barycenter G+ distinct from G-.

We see that when its state changes, each acteon has a probability measure for its co-relations which alternates between around G- and around G+. Between two successive reversals, all acteons have changed of pole. There is a periodicity of two reversals for polar belonging.

First consider the case of Q very isolated (Nx <<Ni) So that internal bunch existst (to a very good approximation). Let us take internal sub-bunches for a single reversal.

The minus intra polar bunch has co-relation of modality -1 : (11) co-relations. It has a complex descriptor $z_1$=-q which is pure real.

The plus intra polar bunch has co-relations of modality +1 : (00) co-relations. It has a complex descriptor $z_2$=+p which is pure real.

The inter pole bunch contains co-relations with modality +j or -j : (10) or (01) co-relations. It has a complex descriptor $z_{12}$=rj which is pure imaginary.

A structure which has electric charge characteristics is a coherent acteon intraverted set which has two poles : a local pole manifesting as a conventional charge, a dispersed pole that will probably interact with the outside to create the effects of the Dirac sea by somewhat different means than those of present theory.

In passing, from this idea filtrates the thought that the dark matter problem could perhaps receive a solution : if half of the matter is under localized form, there is another half formed of dispersed poles. This idea will not be discussed further but is perhaps a way to explore.

We can very schematically illustrate the situation :

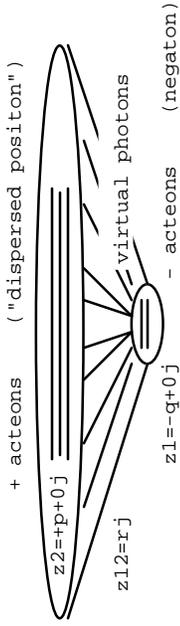

```
+ acteons    ("dispersed positon")
z2=+p+0j
             z1=-q+0j
                        virtual photons
z12=rj                  - acteons   (negaton)
```

We added a mention of virtual photons in this drawing, because the finely divided character of each pole (constituted of acteons with specific polarity) allows for a semi dispersed version of the two charged poles which defines a limited zone of electomagnetic field, that is a photon. The electron configuration is a limiting case of photon where a pole is concentrated and the other completely dispersed so that field fills whole space in between. In this sense an electron is a pure photon set arranged in a special configuration that corresponds to the so called virtual photons ; absorption requires a geometrical adaptation of absorber and absorbed).

Nothing has been said from the charge quantification with e value. It seems natural to relate the charge value to the number of acteons present in an electron pole such that the whole electron is an intraverted structure. That is charge quantum is defines by retroversion scale. This point should be more closely examined, but we shall defer this discussion as some complications are to be expected without invalidating the idea we think though.

relations in the inter bunch link transitions of acteons in the local poles with specific sign, and 1/r probability.

After a state change occurred, acteons of the local pole react as they must do when they are in the dispersed pole, that is a co-relation probability roughly uniform on the whole universe, and thus with very little probability to be in the inter charge bunch. The inter bunch is then exclusively composed of co-relations that interest two acteons of the two local poles, and bring them in the respective dispersed poles.

Between two - poles, the inter bunch has $z_{12}$ real and negative ($\theta_{12}=\pi$), local poles having transitions 1= (-+), and complementary bunches with co-relations numbers $N_1$ and $N_2$ are described by sums of -1 and -j co-relations.

If local poles are sufficiently isolated (relations with the outside is an $\epsilon$ compared to the N), -1 intra local pole co-relations and -j co-relations between the local and the dispersed poles of a same electric strcuture are in nearly equal number, thus $\theta_1$ et $\theta_2$ are close to $5\pi/4$. The relation between two identically charged poles is described with a spinor :

$$\eta_s^A = \frac{1}{\sqrt{N_{12}}} \begin{bmatrix} \sqrt{N_1}e^{j\pi/4} \\ \sqrt{N_2}e^{j\pi/4} \end{bmatrix}$$

For two + local poles, $z_{12}$ is real and positifve ($\theta_{12}=0$), $z_1$ and $z_2$ are composed with co-relation modalities 1 and j.. Assuming a sufficient isolation, $\theta_1$ and $\theta_2$ are close to $\pi/4$, the descriptive spinor is identical to the former.

For two opposed poles, say the first - the second +, the inter bunch has $z_{12}$ pure imaginary and positive (transitions with modality j, $\theta_{12}=\pi/2$), as before, $z_1$ has $\theta_1=5\pi/4$, $z_2$ has $\theta_2=\pi/4$. If charges are inverted, we have $\theta_{12}=-\pi/2$, $\theta_1=\pi/4$, $\theta_2=5\pi/4$.

We thus have a spinor for inter opposite poles relation which reads :

$$\eta_a^A = \frac{1}{\sqrt{N_{12}}} \begin{bmatrix} \sqrt{N_1}e^{j3\pi/4} \\ \sqrt{N_2}e^{-j\pi/4} \end{bmatrix} = \begin{bmatrix} e^{j\pi/2} & 0 \\ 0 & e^{-j\pi/2} \end{bmatrix} \eta_s^A$$

Separations are :

The spinor describing the interrelation of the negaton with the positon is then very special because :

$$\theta=\pi \quad \theta_2=\pi/2 \quad \theta_2=0$$

And the two poles being assembled in the same homogeneous structure, we have reason to believe that :

$$N_1=N_2=N_{12}$$

Whence the descriptive spinor :

$$\eta^A = \begin{bmatrix} e^{j\pi/2} \\ e^{-j\pi/2} \end{bmatrix}$$

And separations :

$$u^t=1 \quad u^z=0 \quad u^x=-1 \quad u^y=0$$

Reversing the roles, the relation of a positon with a negaton is a symetrical situation which gives the conjugated spinor.

## 7) The electrostatic field

Now let us assume that the electric structure is not isolated. Two relations are interesting : relation with the rest of the universe and relation to another electric structure.

Between two identcally signed local charge poles, there is a bunch of co-relations whose characteristics are specific and function of their polarity.

Let us take the case of the quasi point space, and let us suppose that local poles are quasi points, that is their probability measures for co-relations have the 1/r character.

Two charge poles have an inter-bunch with $N_{12}$ co-relations and 1/r proportionality of co-relation probabilities, where r is the z separation between the two poles.

The action co-relation modality between two + charges is 1, between two - charges it is -1, between two opposite charges it is j or -j. Effectively, co-

# G) The quantum level of reality

## 1) Road map to quantum behaviour

Objects of the quantum world are acteon structures. Their apparently strange properties can be explained, for some of them in a very natural way, in the frame previously described.

Those quantum properties have not been discussed until now, because only global and amorphous aspects of action relation were taken in account. Moreover, we did not discuss any dynamical law for acteons' probability measures evolution, and this aspect is fundamental to quantum behaviour intervening in an "inertial like evolution" (the Schrödinger evolution) and in the structural integration of substructures (the measurement process).

Those dynamical aspects have been eluded in the former discussion, where they were important for the inertial movement thought, but will be discussed after quantum behaviour is examined.

A quantum object is an acteon structure which distinguishes by its level of composition (or scale in the universe composition scale). Quantum objects are situated at the retroversion scale, where acteons structures become sufficiently introverted for their internal coherence to show some locality, but their non locality is not completely screened by numerous structural integrations (measurements) which are responsible for the onset of the classical macroscopic behaviour.

a. The measurement process.

What is called a measurement in quantum mechanics is considered here as the integration of an action dynamical structure (for example a photon) in another structure (for example the electronic system of an atom).

The term of integration is used to emphasize the priviledged action relations of the integrated structure in its receptacle, which constitutes an assembly distinct from other structures.

Before a structural integration happens, the candidate to absorption is generally in relation with many other acteon structures, those relations are multiple and diffuse so that no strong implication with a specific structure is present.

$$u_s^t = u_a^t = \frac{N_1 + N_2}{2N_{12}} \quad u_s^z = u_a^z = \frac{N_1 - N_2}{2N_{12}} \quad u_s^x = -u_a^x = \frac{\sqrt{N_1 N_2}}{N_{12}} \quad u_s^y = u_a^y = 0$$

The rank two spinor is :

$$(u_s^\alpha)^{AA'} = \frac{1}{N_{12}} \begin{bmatrix} \frac{1}{2}N_1 & \sqrt{N_1 N_2} \\ \sqrt{N_1 N_2} & \frac{1}{2}N_2 \end{bmatrix} \quad (u_a^\alpha)^{AA'} = \begin{bmatrix} \frac{1}{2}N_1 & -\sqrt{N_1 N_2} \\ -\sqrt{N_1 N_2} & \frac{1}{2}N_2 \end{bmatrix}$$

When placed in the pseudo cristal of test particles as we did for mass, we see that a comparable situation exists. Near a - charge pole, 1 and j co-relations are inexistant ; a - charge pole is thus "1,j isolated". For a + charge pole, it is the contrary, it is "-1,-j isolated". As in the case of mass, the isolation effect decreases in 1/r, which suggests that partial isolation is the origin of the eletric interaction ( as full isolation is the origin of gravitational interaction).

We also see that two identically signed charge poles brought close to each other increase the partial isolation (potential energy increases). For two opposite poles, complementary partial isolations cancel (energy of the system decreases).

In fact, for inexistant co-relation modalities, a situation of selective curvature exists (1,j curvature of - pole or -1,-j curvature of pole +). What complicates the situation is the different reaction of differently signed charge poles in this selective curvature. An opposite pole is affected by a kind of "selective antigravity";

The electrostatic field thus has an origin whose mechanism is analogous to the gravity field, except that the first is an effect of selective curvature, the second being an effect of a curvature from all co-relations isolation.

implicated in actions with other universe objects (amongst them quantum objects).

c. The wave character of quantum objects.

The most difficult point to explain is why quantum objects do posess a periodic behaviour, and why this periodic behaviour is associated to interferences and structural integration probability modulation. The previous paragraph expresses how several potential happenings result from the common multiple relational involvements, but it does not imply any form of periodicity in the relational modalities. The only necessity for a basic form of periodicity arises from the existence of structural stability. But examination of this fundamental point (the whole picture of wave mechanics relies on it with fundamental $E=h\nu$ and $p=hc/\lambda$ formulas) deserves longer developments.

## 2) Structural stability and the quantum pulsations

If an objet is a stable acteon structure, some of its characteristics are stable, that is, some properties of the probability measures for its participating acteons must stay close to a mean value and if they fluctuate, they should oscillate around the mean value. Intuitively, structural stability is the basis for a form of periodicity.

a. The existential thickness.

A remark in the beginning pointed out that along the line representing the continuity of existence for an acteon, one and only one segment represented a state of existence exclusive of the others on the line.

This has the important consequence that an acteon set forming a structure has an existence "slice" in the graph of universe evolution. During its alteration, a structure's existence slides along the continuity lines, but as soon as all acteons in the structure have changed their state, the structure is represented by a new set of existence segments having no common part with the previous existence slice.

The definition of this existence slice is difficult because all action co-relations which do not share an acteon cannot be *a priori* ordered in the universe stochastic process of evolution. Only co-relation issued from an

The measurement process, or structural integration is a kind of "polarization" or "condensation" of the action co-relations by which former diffuse multi object relations change to specifically "oriented" relations defining the clear participation to a definite structure.

It is a form of movement (a non local discontinuous movement), as we defined earlier the movement as a relational change.

Between those structural integration processes (or measurements) an object diffusively evolves its multiple relations in a free non specific way as long as an integrative configuration does not happen and sucks the object into a specific structure for assembly (measures it).

The wave function collapse is the current theoretical picture of this relational change which is indeed a kind of relational contraction, in contrast to the free diffusive evolution which is relationally expansive.

b. The superposition principle

The elementary actions model has non local foundations because acteons are non local entities exterior to space-time their relations built. The non local properties of quantum objects is thus a consequence of acteons non locality.

In addition, the action relations between acteon structures allow for simultaneous co-relation bunches of a structure (the acteons model is a "felt like" relational model allowing non local interleaving of action co-relations).

A quantum structure generally has co-relation bunches with a great number of other structures of common existence. A quantum structure thus has a disposition relatively to many other structures, this multiple disposition has modalities which condition the structural integration process.

The approach to the structural integration and the probability of effective integration (measurement process or interaction) can involve several other objects which are high probabillity integrating candidates. The quantum wave function is a representation of this fact, it is a mathematical description of the compatibility field, ("projection" or "covering" at each space point), of the quantum acteon structure on potentially integrating structures. There is no Schrodinger "cat problem", the state superposition is a picture of the integration potentialities of a unique action co-relation state multiply acteon existence line can be ordered, so that the graph of co-relations restrained to acteons of a structure allows ordering by following co-relation lines.

We can do something though : we define a mean thickness for the existence of a structure S as the mean of state transitions counts insuring all acteons have changed their state at least once in S.

Some special cases can enlighten the idea.

If all acteons in S are pairwise co-related so that all change their state at once, S thickness is one.

If, as is the case in the homogeneous structure, all acteons in S co-relate randomly an uniformly with the others, thickness increases because some acteons will change their state several times before the last of them flips for the first time.

In very structured objects where some structural parts evolve intensely while others "sleep", thickness can have great values.

   b.  Existential co-relation bunches.

The existential thickness of structures is important in the following because it distinguishes co-relation bunches that connect existential slices of common existence (called existential bunches) from all bunches we can take in the evolution graph. The majority of bunches chosen without considering existential thickness will not represent the relational state in an existence stage, but will be regularizing means over an existence interval.

The definition of existential bunches brings other problems though because existential thickness is not the same for all structures. When a thick structure is related to a finer one, an existence slice of the first is related to several of the second. For now we let this question aside.

In the following discussion of quantum behaviour, we shall suppose we have existential bunches, and all structures we are interested in have similar existential thickness.

   c.  Fluctuations of the existential bunches.

When a structure $S_1$ relates to another $S_2$, the different existential co-relation bunches fluctuate because they are realization of a part of the universe stochastic process. Probability measures of the structure acteons can be unchanged, the sampling effect will change the bunch composition.

If we come back to the spinor representation for a co-relation bunch, each existential bunch has a descriptive spinor, and the ordered set of existential bunches corresponds to an ordered set of fluctuating spinors.

This succession of spinors will have some stability properties : if probability measures for co-relations do not change, numbers of co-relation of a given modality will keep stable means and standard deviation. It implies the stability of the z, t and "in XY plane" separations. Those will fluctuate around the mean values.

   d.  Spinor phase diffusion and the Einstein E=hν formula.

The spinor phase is not constrained by the probability measures as the separations above are, so that it is free to diffuse on its domain of variation (only the xy plane separation is constrained).

The phase evolves erratically, but the periodic domain produces a lowest frequency for phase return to a previous value. If the mean phase step between two successive spinor is dθ and the diffusion is a brownian motion on the circle, the displacement for a phase loop is obtained after $N=(2\pi/d\theta)^2$ steps.

We feel that an explanation for the De Broglie pulsation is at hand, but some problems remain : there is a phase spectrum, so that we can wonder where monochromaticity has its origin ; the pulsation does not behave like a clock pulsation in relativity but increases with energy.

The energy has been related to structural isolation, isolation is mainly the result of $N_{12}$ co-relation count being much less than proper structural counts $N_1$ and $N_2$. But the least $N_{12}$ which is at denominator, the greater the phase fluctuation steps are. So that the phase looping period decreases and the phase frequency increases as a consequence of isolation.

   e.  Interference seen as cogent manifestation.

We now have two elements to understand interference. One of them is the phase diffusion and low frequency looping, the other is the measurement process as a structural integration process. We can understand the following fact : if a structurally integrable structure $S_1$ relates to another potentially integrative one $S_2$ with contradictory relational sub-bunches, it cannot be structurally integrated. In fact the contradictory bunches produce no resultant effect which could lead $S_1$ to integrate $S_2$, so that it will not manifest (it will not be present).

As the integration process is stochastic, the contradictory character of several bunches deteriorates smoothly with their phase-amplitude mismatch (the probability amplitude destructive composition).

The usual quantum probability amplitudes for a particle are directly related to the properties of the co-relation bunch (number and modalities counts). In fact the descriptive spinor for a co-relation bunch gives a probability amplitude, and the brownian displacement character of spinor components covers the surfacic character of the co-relation bunch expressed in the complex plane. This surfacic character explains by its covering properties with another surface the probability as the squared modulus of the probability amplitude. The Born interpretation square of $\Psi$ law originates in the brownian character of the co-relation bunch formation process (random sequence of interstructure co-relations).

One more element is missing and some precisions must be given.

At a point of their common existence, two structures have one and only one co-relation bunch. But a co-relation bunch is a global description of all the co-relations, nothing prevents us to divide this unique bunch into several sub-bunches which have their own descriptive spinors (superposed in the spinor for the global bunch).

The question is how parts of the co-relation bunch do become contradictory (get probability amplitude mismatch). The answer resides in the relation of movement and phase, that is in the way the presence of momentum modifies the sequence of existential bunches at the quantum level.